\documentclass[preprint]{aastex62}
\usepackage[USenglish]{babel}
\usepackage[utf8]{inputenc}
\usepackage[T1]{fontenc}
\usepackage{newtxtext,newtxmath}
\usepackage{xspace}

\usepackage{wrapfig}
\usepackage{gensymb}
\usepackage{xfrac}

\graphicspath{{./figs/}}

\newcommand{\hompc}{\,h\,{\rm Mpc}^{-1}}

\begin{document}
\title{Cosmology with the MaunaKea Spectroscopic Explorer}
\author{Will J. Percival}
\affiliation{Centre for Astrophysics, University of Waterloo, Waterloo, Ontario N2L 3G1, Canada }
\affiliation{Department of Physics and Astronomy, University of Waterloo, Waterloo, Ontario N2L 3G1, Canada }
\affiliation{Perimeter Institute for Theoretical Physics, Waterloo, Ontario N2L 2Y5, Canada}

\author{Christophe Y\`eche}
\affiliation{IRFU, CEA, Universit\'e Paris-Saclay, F-91191 Gif-sur-Yvette, France}

\author{Maciej Bilicki}
\affiliation{Center for Theoretical Physics, Polish Academy of Sciences, al.
Lotnik\'{o}w 32/46, 02-668, Warsaw, Poland}

\author{Andreu Font-Ribera}
\affiliation{University College London, Gower St, Kings Cross, London WC1E 6BT}

\author{Nimish P. Hathi}
\affiliation{Space Telescope Science Institute, Baltimore, MD, USA}

\author{Cullan Howlett}
\affiliation{International Centre for Radio Astronomy Research, The University of Western Australia, Crawley, WA 6009, Australia}

\author{Michael J. Hudson}
\affiliation{Department of Physics and Astronomy, University of Waterloo, Waterloo, Ontario N2L 3G1, Canada }
\affiliation{Perimeter Institute for Theoretical Physics, Waterloo, Ontario N2L 2Y5, Canada}

\author{Alan W. McConnachie}
\affiliation{NRC Herzberg Astronomy and Astrophysics, Dominion Astrophysical Observatory, Victoria, BC, Canada}

\author{Faizan Gohar Mohammad}
\affiliation{Centre for Astrophysics, University of Waterloo, Waterloo, Ontario N2L 3G1, Canada }

\author{Jeffrey A. Newman}
\affiliation{University of Pittsburgh and PITT PACC, Pittsburgh, PA ,USA}

\author{Nathalie Palanque-Delabrouille}
\affiliation{IRFU, CEA, Universit\'e Paris-Saclay, F-91191 Gif-sur-Yvette, France}

\author{Carlo Schimd}
\affiliation{Aix Marseille Univ, CNRS, CNES, LAM, Marseille, France}

\author{Andrei Variu}
\affiliation{IRFU, CEA, Universit\'e Paris-Saclay, F-91191 Gif-sur-Yvette, France}

\author{Yuting Wang}
\affiliation{National Astronomy Observatories, Chinese Academy of Science, Beijing, 100101, P.R.China}

\author{Michael J. Wilson}
\affiliation{Lawrence Berkeley National Laboratory, 1 Cyclotron Road, Berkeley, CA 94720, USA}

\correspondingauthor{Will J. Percival, Christophe Y\`eche}
\email{will.percival@uwaterloo.ca, christophe.yeche@cea.fr}

\begin{abstract}
This document summarizes the science cases related to cosmology studies with the  MaunaKea Spectroscopic Explorer (MSE), a highly-multiplexed (4332 fibers), wide FOV (1.5 sq deg), large aperture (11.25 m in diameter), optical/NIR (360nm to 1300nm) facility for obtaining spectroscopy with a resolution $R \sim 2500-4000$.

The MSE High-z Cosmology Survey is designed to probe a large volume of the Universe with a galaxy density sufficient to measure the extremely-large-scale density fluctuations required to explore primordial non-Gaussianity and therefore inflation. We expect a measurement of the level of non-Gaussianity as parameterized by the local parameter $f_{NL}$ to a precision $\sigma(f_{NL}) = 1.8$.

Combining the MSE High-z Cosmology Survey data with data from a next generation CMB stage 4 experiment and existing DESI data will provide the first $5\sigma$ confirmation of the neutrino mass hierarchy from astronomical observations. Only combining the data from the MSE High-z Cosmology Survey together with Planck provides a $4\sigma$ neutrino mass measurement.

In addition, the Baryonic Acoustic Oscillations (BAO) observed within the sample will provide measurements of the distance-redshift relationship in six different redshift bins between $z=1.6$ and 4.0, each with an accuracy of  $\sim0.6$\%. These high-redshift measurements will provide a probe of the Dark Matter dominated era and test exotic models where Dark Energy properties vary at high redshift. The simultaneous measurements of Redshift Space Distortions (RSD) at redshifts where Dark Energy has not yet become important directly constrain the amplitude of the fluctuations parameterized by $\sigma_8$, at a level ranging from $1.9\%$ to $3.6\%$ for the same redshift bins.

The proposed survey covers 10,000 ${\rm deg}^2$, measuring redshifts for three classes of target objects: Emission Line Galaxies (ELGs) with $1.6<z<2.4$, Lyman Break Galaxies (LBGs) with $2.4<z<4.0$, and quasars $2.1<z<3.5$. The ELGs and LBGs will be used as direct tracers of the underlying density field, while the  Lyman-$\alpha$ (${\rm Ly}\alpha$) forests in the quasar spectra will be used to probe structure along their lines of sight.

Exposures of duration 1,800\,sec  will guarantee a redshift determination efficiency of 90\% for ELGS and at least 50\% for  LBGs. After optimization of the tiling,  the survey will be comprised of 8,000 pointings, and will take 4,000\,hours on target in total, representing 100 nights per year for a 5-year MSE program. 

In addition to the high-z survey, three ideas for additional projects of cosmological interest are proposed: a deep survey for LSST photometric redshift training, pointed observations of galaxy clusters to $z =1$, and an IFU-based peculiar velocity survey.

\end{abstract}

\tableofcontents

\section{A large-volume survey of high-redshift galaxies}

\subsection{Background Cosmology}

Observations over the last 50 years have provided a tremendous insight into the Universe, with many of the key parameters including the age of the Universe and the split of energy-density components determined with high precision (e.g. \citealt{Planck2018}). The standard cosmological model, also known as $\Lambda$CDM, postulates that the late-time evolution of the Universe is driven by Dark Energy in the form of a cosmological constant ($\Lambda$) \citep{Riess98,Perlmutter99}.  Although this model is a tremendous success, matching measurements of ever-increasing precision, this has thrown up some big questions: the origin of a $\Lambda$ term is difficult to understand physically, suggesting that the $\Lambda$CDM might be an approximation to a more complicated theory (e.g. \citealt{mortonson14}. Our knowledge of the very early Universe is also limited, except for the evidence that a period of rapid acceleration (called inflation) is needed to solve a number of problems with the standard cosmological model; we do not know exactly what drives this inflation, however (e.g. \citealt{liddle99}). We also do not know how structure growth affects the large-scale cosmological model (a concept often called back-reaction, e.g. \citealt{buchert12}). Finally, there are a number of unknown physical parameters, such as the summed neutrino particle mass, whose influence spans the fields of cosmology and particle physics (e.g. \citealt{Lesgourgues12}).

At least two neutrino species are known to have non-negligible mass thanks to flavor oscillation experiments \citep{sudbury01,Super-Kamiokande06}. However, current observations are consistent with many neutrino mass models, and determining the absolute mass scale is an obvious goal in the field of particle physics. It is the target of terrestrial experiments such as the searches for neutrinoless double beta decay \citep{Delloro16} or tritium beta decay experiments \citep{Otten08}, but can also be measured through astronomical observations \citep{Lesgourgues12}. The best current constraints on the summed neutrino mass are $\sum m_\nu < 0.12\; {\rm eV}$ (95\% confidence), combining low-redshift BAO data with the 2018 CMB data from Planck \citep{Planck2018}. The normal hierarchy with one particle of negligible mass has $\sum m_\nu = 0.057\; {\rm eV}$, while the inverted hierarchy with one negligible mass neutrino has $\sum m_\nu = 0.097\; {\rm eV}$. Thus, for example, we need to measure the neutrino mass with an error of $\sigma =0.008\; {\rm eV}$  in order to rule out the inverted hierarchy at $5\sigma$ if neutrino masses are distributed in the normal hierarchy and $\sum m_\nu =  0.057\; {\rm eV}$. As we will see below, the MSE will be a vital component in enabling such a measurement, which is achievable when the MSE data is combined with other available cosmological data.

Measuring the level of primordial non-Gaussianity is one of the most powerful ways to test inflation, and the high-energy early Universe (e.g. \citealt{Bartolo04}). The simplest slow-roll models of inflation predict the generation of primordial fluctuations that are almost Gaussian distributed, with only a tiny deviation from Gaussianity. If we consider fluctuations in the potential arising after inflation, then one can denote the portion of the potential that can be described as a Gaussian random field as $\varphi$ and assume that the level of primordial non-Gaussianity is a local function of the potential. To 2nd order, this approach yields $ \Phi =  \varphi + f_{NL}(  \varphi^2 - \langle\varphi^2\rangle)$, where $f_{NL}$ is the parameter that we want to measure. A local designator is often ascribed to the parameter, as defined in this way, but for brevity we do not do this. The best current constraints on  $f_{NL}$ come from Planck, who in 2015 (the 2018 update has yet to be published) used Bispectrum measurements of the CMB to measure $f_{NL}=2.5 \pm 5.7$, consistent with no signal \citep{Planck2015PNG}. A detection of $f_{NL} $>1 would rule out the class of slow-role models. The proposed cosmology survey with MSE will provide a measurement with error $\pm 1.8$ from the power spectrum alone. Including constraints from the Bispectrum would further improve these predictions \cite{Karagiannis18}.

Although the standard cosmological model has been an incredible success, there are tensions between measurements at the $2-3\sigma$ level. The most significant of these (at $3.5\sigma$) is the mismatch between Hubble parameter measurements made locally with supernovae \citep{Riess18}, and those obtained from the combination of galaxy survey and Cosmic Microwave Background (CMB) measurements \citep{Planck2018}. In addition, locally measured structure growth rates from weak lensing \citep{KiDS_WL2018,DES_WL2008} differ from the $\Lambda$CDM extrapolation of CMB measurements, while RSD measurements are more in agreement \citep{Alam2016}. In the past, $2-3\sigma$ tensions between data sets such as these have been revealed as unknown systematic experimental errors, and so it pays to be skeptical. However, these observations might be a hint of new physics so, given the physical issues with the standard model, there are many reasons to investigate further. The proposed MSE survey will push consistency tests to redshifts where there are no current constraints, offering tremendous discovery space.

For the Maunakea Spectroscopic Explorer, we propose a large-volume galaxy survey designed to probe the inflationary Universe through measurement of primordial non-Gaussianity, and to measure the summed neutrino mass both to physically interesting levels. A by-product of this survey will be more standard measurements of the distance-redshift relation and rate of structure growth, testing the standard model over a redshift range not studied prior to this survey. These measurements will test the $\Lambda$CDM model in ways (particularly the redshift range) not previously tested, offering the potential to discover deviations.

\subsection{Galaxy Redshift Surveys}

Galaxy surveys provide several complementary mechanisms by which the cosmological measurements described above can be made. 

Massive neutrinos alter the observed distribution of galaxies by changing the matter-radiation equality scale compared with the same cosmological model without massive neutrinos: as they are still relativistic at decoupling they hence effectively act as radiation instead of matter around the time of equality. This produces an enhancement of small-scale perturbations in the CMB, especially near the first acoustic peak, and slightly alters the linear matter power spectrum as traced by galaxies. Their effect on structure growth in the matter-dominated era leaves a clearer imprint on galaxy redshift survey measurements due to their relativistic to non-relativistic transition. Massive neutrinos suppress growth on scales below the Hubble radius at the time of the non-relativistic transition, because free-streaming leaves the density contrast for neutrinos lower than that of CDM and they can only slowly catch up through growth. On scales larger than the Hubble radius at the transition time, fluctuations in the distribution of neutrinos are of the same order as those in CDM, as free-streaming of neutrinos out of overdensities is ineffective on these scales and so growth is not as suppressed. A review of these effects is given in \citet{Lesgourgues12}. As a result, the summed neutrino mass can be measured from the scale-dependent growth of matter fluctuations as traced by galaxies. This is the primary mechanism that MSE will exploit in order to measure the summed neutrino mass.

Primordial non-Gaussianity alters the expected power spectrum of galaxies in a complementary way to the effects of massive neutrinos. The change in the relative number of high mass haloes that the non-Gaussianity creates alters the expected mass function of dark matter halos in such a way as to produce a scale dependent bias for galaxies that diverges as $1/k^2$, where $k$ is the Fourier wave number of the clustering modes \citep{Dalal08}. The best $f_{NL}$ measurements to date from a spectroscopic survey used the Baryon Oscillation Spectroscopic Survey (BOSS), finding $f_{NL}$ consistent with zero and an error of order $\pm100$ \citep{Ross13}. This will soon be supplanted by the extended-BOSS survey, which is expected to provide constraints of order $\pm 10$ \citep{eboss-predictions}. As this scale-dependent bias arises on very large scales, we need large survey volumes in order to beat down sample/cosmic variance. The eBOSS survey achieves this by observing quasars out to redshift 2.2 over of order 2000 ${\rm deg}^2$ (for the DR14 sample; the final sample should total 4500 ${\rm deg}^2$). By pushing to a larger sky area and to higher redshifts, MSE can significantly improve on these measurements. In this document we adopt a conservative strategy and only consider $f_{NL}$ constraints coming from the power spectrum. Measurements made from the Bispectrum have the potential to improve these measurements by a factor $\sim1.5$ (see Table 9 of \citealt{Karagiannis18}) for local $f_{NL}$ measurements, and a larger factor for other types of primordial non-Gaussianity to which the Bispectrum is more sensitive than the Power Spectrum. However these measurements come at the cost of an increased reliance on being able to model non-linear effects, and we do not present any explicit predictions here.

More standard cosmological measurements come from features in the clustering pattern of galaxies arising due to Baryon Acoustic Oscillations (BAO). BAO are caused by acoustic waves in the early Universe, and act as a standard ruler of fixed comoving length, whose apparent size when observed can be used to constrain the distance-redshift relationship and the geometry of the Universe. Information about the growth rate of large-scale structure is obtained through Redshift-Space Distortions (RSD). RSD arise because we do not observe true galaxy positions, but instead infer distances from measured redshifts, which include coherent flows due to the growth of structure. A review of the physics of BAO and RSD is given in \citet{Percival13}. The combination of geometrical and structure-growth measurements has incredible power to constrain theories of acceleration based on modifications to GR (e.g. \citealt{Joyce16}).  If GR is correct, the growth of large-scale-structure fluctuations can be predicted directly from the expansion history; deviations from this relationship indicate that new physics is needed.  Conversely, non-GR theories of gravity cannot be easily tuned to match both the expansion rate and structure growth rate simultaneously. 

In addition to the probes discussed above, once a galaxy redshift survey has been produced it acts as a powerful resource for other cosmological measurements and tests, such as those made using the bispectrum \citep{gilmarin15} or from identifying clusters of galaxies or voids \citep{nadathur16,Mao17}. One also obtains significant information by cross-correlating galaxy surveys with other data. For MSE, one interesting avenue will be the cross-correlation of the galaxy survey with CMB data. In particular, the MSE High-z Cosmology Survey will include coverage over $2<z<3$, which corresponds to the peak of the CMB lensing effectiveness \citep{Planck18-lensing}. 

Although these science interests are not the focus of the design of the survey described here, it is important to remember that this survey will offer significant legacy both within the field of cosmology and beyond.

\subsection{The MSE High-z Cosmology Survey}

The MSE High-z Cosmology Survey is designed to probe a large volume of the Universe with a density of galaxies sufficient to measure the extremely-large-scale density fluctuations required to explore primordial non-Gaussianity. By pushing to high redshifts, the growth of structure in the Universe is closer to linear dynamics on average, potentially simplifying the modeling, although the strong biasing of the galaxy samples to be observed will limit the extent of this improvement.

The proposed strategy is not optimized to make BAO and RSD measurements at high redshift, as Dark Energy does not dominate the cosmological energy budget here; however, measurements of these quantities still form a secondary goal of the survey. RSD measurements at redshifts where Dark Energy has not yet become important, directly constrain the amplitude of the fluctuations parameterised by $\sigma_8$. BAO will measure the angular diameter distance to high redshift, providing a probe of the Dark Matter-dominated era and tests of exotic models where Dark Energy properties vary at high redshift. 

Our proposed survey covers 10,000 ${\rm deg}^2$, measuring redshifts for three classes of target objects: Emission Line Galaxies (ELGs), Lyman Break Galaxies (LBGs), and quasars. The ELGs and LBGs will be used as direct tracers of the underlying density field, while the the Lyman-$\alpha$ (${\rm Ly}\alpha$) forests of the quasars will be used to probe structure along their lines of sight. Details of our proposed target selection and estimated redshift efficiencies are given below. Fig.~\ref{fig:cosmo_survey} shows how the proposed MSE survey compares to other galaxy redshift surveys. 

We propose to undertake exposures of duration 1,800\,sec. Each exposure covers 1.52 square degrees and we expect to pack the observations onto the sky to cover the maximum area possible without gaps, such that each exposure provides approximately 1.25 square degrees of extra coverage. Thus the survey will be comprised of 8,000 pointings, and will take 4,000\,hours on target in total. With the expected MSE efficiency of 2,336\,hours on target per year, the survey will take approximately 1.7\,years of telescope time in total to complete. As we will only use Dark-Time observations, the survey will need to be spread over a number of years, for example taking $\sim$100 nights per year over 5 years, interspersed with other programs. 

\begin{figure}[htbp]
\begin{center}
\includegraphics[angle=0,  clip, width=16cm]{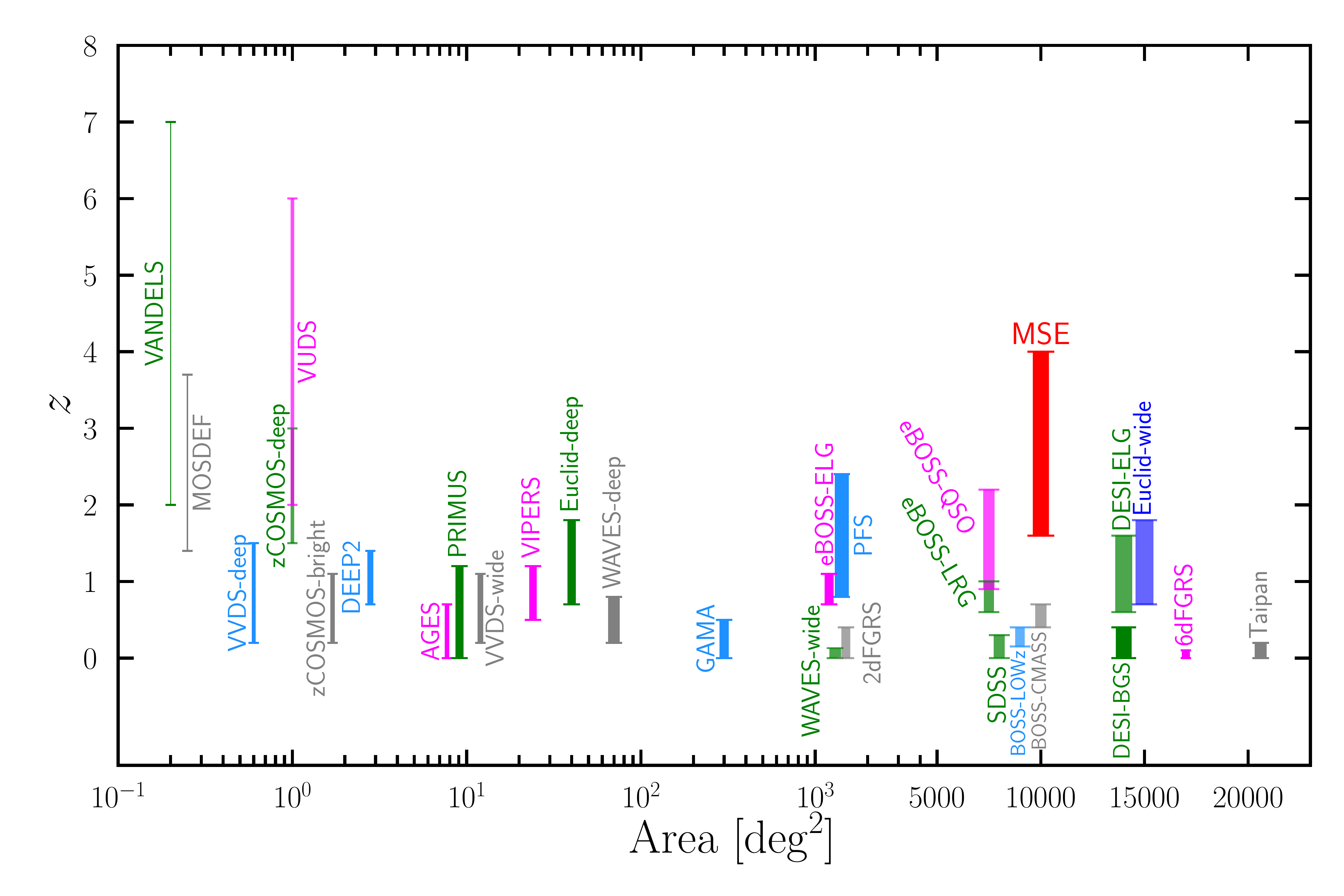}
\caption{\it Recent galaxy redshift surveys as a function of their area and redshift range, compared with the proposed MSE survey. The thickness of each bar is proportional to the total number of galaxies. Notice the transition from logarithmic to linear scale on x-axis at $5000\,\mathrm{deg}^2$.}
\label{fig:cosmo_survey}
\end{center}
\end{figure}

The area to be covered by the survey is limited by the availability of imaging data from which to target, while the number of galaxies to be targeted and redshifts obtained will be limited by the number of fibers and the total exposure time available for the survey. Having more fibers on MSE would improve the scientific return, although the return would diminish with each fibre added as we will always preferentially target the galaxies most likely to provide cosmological information. Similarly, increasing the exposure times would improve the efficiency of successful redshift measurement, and would be particularly useful given an increase in the number of fibres. For hardware reasons, adding more fibres may only be possible with an increased Field of View (FoV). Note that increasing the FoV while keeping the number of fibres fixed does not significantly help given that we wish to increase the density of observations rather than increase the survey area within a fixed total observing time. Techniques developed for the Dark Energy Spectroscopic Instrument (DESI) will be used to mitigate selection effects given the patrol radii of individual fibres \citep{Burden17, Pinol17, Bianchi17, Percival17}, and thus we do not expect any problems with accurately measuring clustering due to sampling targets for observation.

\subsection{Detailed predictions for targeting and exposure times}

We will use three tracers:
\begin{itemize}
\item[]  1.6<z<2.4: Emission Line Galaxies (ELGs)
\item[]  2.4<z<4.0: Lyman Break Galaxies (LBGs) and ${\rm Ly}\alpha$ emitters (LAEs)
\item[]  2.1<z<4.0: ${\rm Ly}\alpha$ forests of QSOs (${\rm Ly}\alpha$ QSOs)
\end{itemize}

We want to optimize the tiling for the survey outlined above. With 3250 low-res fibres per 1.25\,deg$^2$ gained for the survey, we have a budget of 2600 fibres / deg$^2$ on average. Assuming that $\sim10\%$ of the fibers are used for calibration (sky and reference star fibers), the available density for science targets is $\sim2340$ fiber per deg$^2$. We will share the same pointings between the three tracers (ELGs, LBGs and ${\rm Ly}\alpha$ QSOs); hence all will receive the same exposure time. The fiber assignment priorities will be as follows: first the ELGs, then ${\rm Ly}\alpha$ QSOs and finally the LBGs.  

First, we observe observe the ELGs with a density adequate for BAO measurements, i.e, $nP(k=0.1{\rm h.Mpc^{-1}})=1$.  Assuming a redshift efficiency of order 90\%, the density required is 600 ${\rm deg}^{-2}$ targets. To achieve such an efficiency for ELGs with $r<24$, as explained below we calculate that this requires an exposure time of 1800s. In addition this exposure time allows us to observe the ${\rm Ly}\alpha$ forest in QSOs with $r<24$ or even $r<24.5$, with a SNR per resolution element of the order of 2-3 in the forest, which is optimal for studies of the cross-correlation between the HI absorption and the positions of the other tracers. Assuming the quasar luminosity function of \cite{Palanque2013}  we can expect there to be 150 to 170 ${\rm deg}^{-2}$  quasars with $z>2.1$. The rest of the fibres will be filled with LBGs at the level of approximately 1400 ${\rm deg}^{-2}$  targets. As the exposure time of 1800s is driven by getting a high redshift efficiency for the ELGs, as estimated below we expect for the LBGs a lower redshift efficiency at the order of 50\%. As a result the LBG density is not optimized for BAO measurements but it is still sufficient for the measurement of the primordial non-Gaussianity, i.e, $f_{NL}$. Note that the combination of all three samples together with the sky fibres, fits within the budget for our survey set at 2600 fibres/deg$^2$, with a small amount of margin remaining.

\subsubsection{Emission Line Galaxies}

ELGs are being intensively used to trace the matter within the redshift range 0.6<z<1.6 in the recent survey eBOSS and the upcoming DESI survey. These galaxies are characterized by high star formation rates, and therefore exhibit strong emission lines from ionized H~II regions around massive stars, as well as SEDs with a relatively blue continuum, thanks to which they be selected from optical $ugriz$-band photometric surveys such as LSST in the southern hemisphere or CFIS+Unions in the northern hemisphere. The prominent [OII] (3727 Angstrom) doublet in ELG spectra consists of a pair of emission lines separated in rest-frame wavelength by 2.78 Angstroms.  The wavelength separation of the doublet provides a unique signature if spectral resolution is sufficiently high, allowing definitive line identification and secure redshift measurements. 

In contrast to DESI, the NIR coverage of the low resolution (LR) spectrograph (940nm<$\lambda$<1320nm) allows us to determine ELG redshifts up to $z=2.5$, opening a new research window for RSD and BAO studies.  Considering an $r$-band magnitude limit of $r<24$, the next generation of photometric surveys such as LSST will provide a photometry deep enough to select ELGs based on their optical colors (u-r, g-r and r-z). A u-r or g-r color cut combined with a r-z color cut is ideal for selecting galaxies in the desired redshift range, $1.6<z<2.4$.
The strategy is to measure the UV excess which has a strong correlation with the star-formation activity of ELGs by limiting the u-r or g-r colors. In addition, as the OII line and the Balmer break are redshifted beyond the r-band and the Lyman break is not in the g band, the ELGs are expected to be blue in g-r color. Following the strategy proposed in the PFS proposal~\citep{Takada2014} for a similar ELG survey, a pure optical bands is sufficient to achieve the required ELG density. This selection was  successfully tested over the COSMOS field. However it may suffer from a small contamination by low-z galaxies. 

To avoid this possible contamination, as the ELGs are redshifted compared to DESI ELGs, a selection based on NIR, can improve its purity.  The future LSST project with its Y band, can provide such a NIR photometry. Currently, we can consider KiDS+VIKING in the south hemisphere and HSC in the north hemisphere, as test-beds for developing an optical and NIR approach.  

 Therefore,  it is reasonable to assume that we will be able to select a minimum of 600 ${\rm deg}^{-2}$ ELGs in the desired redshift range. Based on the PFS study~\citep{Takada2014} with a smaller 8m telescope, which will employ a spectrograph with a identical resolution in NIR and same 1800s exposure time, we can expect a redshift efficiency of the order of 90\% for ELGs.

\subsubsection{Lyman Break Galaxies}

The LBGs are strongly star-forming galaxies with blue spectra longward of the Lyman break (at 912 Angstroms). Shortwards of the break almost all light is absorbed by neutral HI along the line of sight.  At $2.4<z<3.5$ they are thus selected by having a very red u-g color combined with very blue g-r color. For more distant quasars, $z \sim 4$, we can use a similar strategy based on g-r and r-i colors. For more than a decade, these selections using the Lyman break color techniques have been used extensively. The primary uncertainty for this tracer is on our ability to determine the redshift for samples with r-band magnitude limit $r<24$ or even $r<24.5$.

Recent observations with MUSE on the VLT~\citep{Caruana2018} have shown that at the redshifts of interest more than half of LBGs exhibit a detectable ${\rm Ly}\alpha$ emission line, which can help to facilitate the determination of  redshifts, as we demonstrate below. For the LBGs without ${\rm Ly}\alpha$ emission, the redshift can be determined thanks to the position of the ${\rm Ly}\alpha$ and ${\rm Ly}\beta$ absorption features and a number of interstellar medium absorption lines.

Given an exposure time of 1800s, we wish to estimate the efficiency to determine redshifts for LBGs at 0.1\% precision $(0.001\cdot(1+z))$. To compute the redshift efficiency we adopt a conservative approach by using two different sets of templates for LBG spectra; one is used to create ''simulated spectra'' and the other is used to determine the redshifts of the simulated spectra.

The three templates from Fig.7 of \cite{Hathi2016} are implemented in the official MSE Exposure Time Calculator (ETC, http://etc-dev.cfht.hawaii.edu/mse/) and thus we obtain the SNR per resolution element for each of them, given a set of parameters (sky brightness, airmass, exposure time, seeing, source type, magnitude of the source and its redshift). In order to obtain the simulated spectra we just redshift the templates and then add gaussian noise given by the SNR.  

These spectra are given to a redshift finder, called PandoraEZ~\citep{Garilli2010}, which measured their redshifts using the seven spectrum templates from  Fig. 15 of \cite{Bielby2011} and Fig. 14-15 of \cite{Bielby2013}. Then we count the number of LBGs having a redshift which fulfills the condition that $|z_{meas}-z_{real}|<0.001\cdot(1+z_{real})$. In each set of templates, the difference between the spectra is the strength of the ${\rm Ly}\alpha$ emission line, expressed through the value of the Equivalent Width (EW). 

\begin{figure}[htbp]
\begin{center}
\includegraphics[angle=0,  clip, width=18cm]{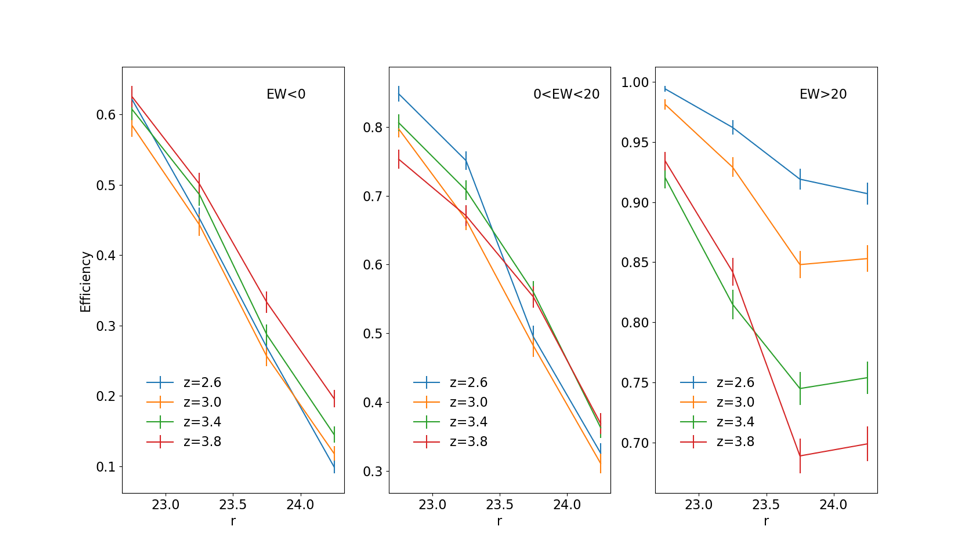}
\caption{\it Redshift efficiency as function of  r magnitude for the 3 templates from \cite{Hathi2016} and for 4 redshift bins. The efficiency does depend strongly on the strength of the ${\rm Ly}\alpha$ emission line and on continuum magnitude.}
\label{fig:cosmo_efficiency}
\end{center}
\end{figure}

In Fig.\ref{fig:cosmo_efficiency}, we present the redshift efficiency as a function of the r magnitude for the three different templates ($EW<0$, $0<EW<20$, and $EW>20$, the last case corresponding to ${\rm Ly}\alpha$ emitters (LAE)) and for 4 redshift bins. One can see that the redshift efficiency is higher for the LBGs with stronger ${\rm Ly}\alpha$  emission line, demonstrating that the determination of the redshift is facilitated by the presence of this emission line. The next step is to make an average over the three templates assuming the ${\rm Ly}\alpha$ emitting fraction of LBGs from \cite{Caruana2018} (i.e. 40\% of LBGs having $EW<0$, 30\% having 0<EW<20 and 30\% having $EW>20$).

Using the luminosity function of LBGs from~\cite{Steidel99}, the average redshift efficiency and the average density of LBGs with a good redshift measurement is computed. For z=3 and $r<24.0$ we obtain an efficiency of 56\% resulting in a density of 218 LBGs with good z per ${\rm deg}^2$ , while for $r<24.5$ the efficiency was 44\% with a number density of 481 LBGs with good z per ${\rm deg}^2$. Consequently, we demonstrate that we have enough LBGs and we have at least 50\% redshift efficiency for $r<24$ to ensure the required precision used by the cosmological forecasts.

\subsubsection{Ly-$\alpha$ forest}

During the last few years, the BOSS collaboration has shown that it is possible to study the large scale structure of the Universe using the ${\rm Ly}\alpha$ forest of absorption features in the spectra of high-z quasars. In their last analysis, the combination of ${\rm Ly}\alpha$ auto-correlation and its cross-correlation with quasars provided a BAO measurement at z=2.4 with a $2\%$ accuracy. 

While low-z galaxy BAO measurements are dominated by sample/cosmic variance, this is not the case for high-z ${\rm Ly}\alpha$ BAO, where the error bars can be dramatically reduced by increasing the density of lines of sight. MSE can increase the density by a factor of 10, bringing ${\rm Ly}\alpha$ BAO analyses to sub-percent precision. MSE will dedicate $\sim170$ ${\rm deg}^{-2}$  fibers to ${\rm Ly}\alpha$ quasar targets. Assuming a targeting efficiency of $90\%$ for $r<24$ targets, this translates into 150 ${\rm deg}^{-2}$ forests, close to the cosmic variance limit of $nP=1$ (see \cite{McQuinn2011} for a discussion of ${\rm Ly}\alpha$ forecasts). In addition to a color selection similar to what is used in DESI, for the footprint overlapping LSST, the quasar can be selected by their intrinsic variability with LSST.

In addition, cross-correlations between the ${\rm Ly}\alpha$ forest and the other galaxy tracers (ELGs and LBGs) will allow multiple internal cross-checks. The different catalogs will be affected by very different systematics, and consistency between the different measurements will make our results more robust.

\subsection{Cosmological measurements}

Given the uncertainties in targeting, we adopt rounded numbers for our baseline predictions, assuming a 10,000 ${\rm deg}^2$ survey consisting of two samples:

\begin{itemize}
\item  ELG sample: 1.6<z<2.4, 5.4M galaxies (with a 540 ${\rm deg}^{-2}$  density corresponding to 600 ${\rm deg}^{-2}$ targets)
\item  LBG sample: 2.4<z<4, 7.0M galaxies (with a 700 ${\rm deg}^{-2}$  density corresponding to  1400 ${\rm deg}^{-2}$ targets)
\end{itemize}

For both ELGs and LBGs, we assume a large-scale structure bias $b(z) = G(0)/G(z)$, where G is the linear growth rate of overdensities in the matter. This assumption was adopted for eBOSS ELGs \citep{Zhao2016} and is based on \cite{Comparat2013a} and \cite{Comparat2013b}. For LBGs, this matches the results of \cite{Jose2013}; empirically, the overall clustering of luminous galaxies on linear scales changes little with redshift, even as the underlying dark matter overdensities change greatly in amplitude.

We predict the constraints on the cosmological parameters obtainable with the MSE using the Fisher Information Matrix. This quantifies the amount of information available about a particular parameter within some observable and accounts for the first-order correlations between different parameters. In our case the observable of interest is the two-point clustering (the power spectrum) of ELGs and LBGs measured with the MSE, whilst the cosmological parameters are the standard parameters of the $\Lambda$CDM cosmological model, the sum of neutrino masses, and $f_{NL}$ as a measure of primordial non-Gaussianity. The inverse of the Fisher matrix provides a lower limit on the expected variance of these parameters and can be calculated using only a model of the galaxy power spectrum that we might expect to measure with the MSE and the baseline numbers for the ELG and LBG samples given above.

Hence, for our given baseline survey, we can provide simple estimates for how well a high-redshift survey of galaxies undertaken with the MSE will allow us to constrain both primordial non-Gaussianity and the sum of neutrino masses. This is done in the following sections. In both cases we anticipate being able to measure the galaxy power spectrum across a range of scales between $k_{min}=2\pi/V^{1/3}$ and $k_{max}=0.2h$/Mpc where $V$ is the comoving volume of the Universe covered by our proposed survey; for a 10,000 ${\rm deg}^2$ sky area between 1.6<z<4.0, $V\simeq90({\rm Gpc}/h)^3$.  We choose our value of $k_{max}$ based on what is achievable with our current state-of-the-art theoretical modelling of the non-linear clustering of galaxies. However, in the era of MSE it is likely we will be able to model even further into the non-linear regime and so also present some scenarios with a slightly higher $k_{max}=0.25h$/Mpc. In this scenario we find that our proposed MSE survey will produce tighter constraints on the sum of neutrino masses than any other galaxy survey and when combined with other data (namely from the highly complementary DESI survey and future CMB experiments) will offer the first $5\sigma$ confirmation of the neutrino mass hierarchy, a fundamental measurement for particle physics. 

\subsubsection{Predictions for neutrino mass}\label{sec:neutrinos}

For ease of comparison with other upcoming surveys, we use the method of \cite{Font-Ribera2014} for our neutrino mass error estimates, which was also adopted by \cite{DESI2016}. We consider an MSE survey where we measure the anisotropic power spectrum of ELGs and LBGs in narrow redshift bins of $\Delta z = 0.1$ between the $k_{min}$ and $k_{max}$ given previously. In each redshift bin we marginalise over the unknown effects of galaxy bias. Our model for the clustering of the ELGs and LBGs is sensitive to the sum of neutrino masses in a number of ways. Different neutrino masses change the expansion history of the Universe, changing the length scales associated with a measurement of the power spectrum. Neutrinos also affect the way in which structures grow, adding a scale dependence to an otherwise scale-independent growth rate. Different neutrino masses will also change the shape of the CDM+baryon power spectrum and its normalisation. Hence, a measurement of the clustering of ELGs and LBGs using the MSE is rich in information regarding the neutrino masses.

For the full MSE ELG+LBG sample from $z=1.6-4.0$ and combining with \textit{Planck} constraints on the standard $\Lambda$CDM parameters, we obtain an error on the sum of neutrino masses of $0.018$\,eV. This is better than any other current or planned survey and would enable a $3\sigma$ constraint on the sum of neutrino masses in either hierarchy. For comparison, the DESI survey between $z=0.0-1.6$ forecasts an error of $0.020$\,eV, however realising such a measurement from DESI will require careful cross-correlation of several overlapping galaxy samples. Our proposed survey is somewhat less complex with samples separated in redshift, and benefits from the increased cosmological volume available with the MSE.

In addition to its ability as a stand-alone survey, a strength of our proposed survey is in its complementarity with other planned projects. In the era of MSE, the DESI data will be publicly available and can be combined with the survey proposed here. The two surveys can be combined easily given their lack of overlap, but together will form an enormous survey stretching from $z=0.0-4.0$, with a combined volume of over $130({\rm Gpc}/h)^3$. We forecast a combined error on the sum of neutrino masses, including current CMB data from \textit{Planck}, of $0.013$\,eV ($0.012$\,eV), for $k_{max}=0.2\hompc (0.25\hompc)$, which corresponds to a $4\sigma$ constraint on the sum of neutrino masses, \textit{and} a $3\sigma$ detection of the difference between hierarchies. 

Finally, adding in information from potential future CMB experiments (i.e., a `Stage 4' CMB experiment following \citealt{CMBS42016})\footnote{Constraints on $\Lambda$CDM parameters for such an experiment provided by Ren\'{e}e Hlo\v{z}ek (private communication) using the \textsc{OxFish} code \citep{Allison2015} after representative foreground cleaning.} will enable the constraint on the sum of neutrino masses of $0.008-0.009$eV depending on the value of $k_{max}$ used. This will be the first $5\sigma$ constraint on the neutrino mass estimate in either hierarchy, \textit{and} of the difference between hierarchies. Such a measurement is only possible with the inclusion of data from the MSE cosmology survey of ELGs and LBGs.

\subsubsection{Predictions for $f_{NL}$} \label{sec:fnl}
In addition to neutrino mass estimates, the extremely large cosmological volume available to us with the baseline MSE cosmological survey compared to other planned surveys will enable exquisite constraints on the amount of primordial non-Gaussianity in the early universe and allow us to rule out a number of inflationary models. Using the same Fisher matrix method as before, we can predict the error on the primordial non-Gaussianity parameter $f_{NL}$ that we can achieve with our baseline survey.

Unlike neutrinos, the effect of primordial non-Gaussianity is to add a scale-dependence to the large scale linear bias of galaxies with respect to the underlying dark matter. The unique $1/k^{2}$ dependence of the galaxy bias introduced by primordial non-Gaussianity also means that the lower values of $k$ that can be reliably measured (i.e., the larger the cosmological `baseline' available to us), the stronger constraints we can obtain. The fact that a cosmology survey with MSE will allow us to probe higher redshifts and larger cosmological volumes than other surveys is a strength that will result in superior constraints

To demonstrate this, we take the approach used for the eBOSS survey \citep{Zhao2016}, and again envision measuring the clustering of ELGs and LBGs observed by the MSE in narrow redshift bins. For primordial non-Gaussianity forecasts, we take the bias and $f_{NL}$ parameters as free parameters, and report the precision of $f_{NL}$ with the bias parameter marginalised over. The minimum wave-vector we used in the Fisher matrix calculation is determined by the volume of the survey, i.e., $k_{\rm min}=2\pi/V^{1/3}$. We found that MSE ELGs can constrain primordial non-Gaussianity of the local form to a precision $\sigma(f_{NL}) = 4.1$. MSE LBGs at higher redshifts can reach a tighter constraining precision of $\sigma(f_{NL}) = 2.0$. The individual QSOs that give rise to our proposed Ly$\alpha$ forest sample can also be used as tracers of the density field and reach a precision of $\sigma(f_{NL}) = 5.7$, which itself is already at the level of current CMB constraints. 

Overall, the combination of MSE ELGs and LBGs will provide a constraint on primordial non-Gaussianity (the local ansatz) to a precision $\sigma(f_{NL}) = 1.8$. This will be further improved by including the QSOs measured in the same volume; measurements of $f_{NL}$ have been shown to be greatly improved by the use of multiple overlapping tracers of the same density field with different galaxy bias \citep{Seljak2009}. This measurement is better than the constraint from CMB temperature and polarisation data of Planck 2015, reaching $\sigma(f_{NL}) = 5.7$ \citep{Planck2015PNG}, and also more stringent than the predictions of $\sigma(f_{NL}) = 15$ by ongoing eBOSS survey \citep{Zhao2016}, $\sigma(f_{NL}) = 5$ by DESI \citep{DESI2016}, and $\sigma(f_{NL}) = 11$ by PFS \citep{Takada2014}.  

\subsubsection{Predictions for BAO \& RSD}

We present the predictions on BAO RSD precision by MSE in Table \ref{tab:baorsdresult}. For BAO forecasts, we follow the method of \cite{Seo2007}. Using both ELGs and LBGs, we expect to obtain 6 measurements in different redshift bins, each with accuracy $\sim0.6$\%. Binned in a slightly different way, we would obtain approximately 17 1\% BAO measurements. These measurements would provide an exquisite determination of the distance-redshift relationship during the matter dominate era over the redshift range $1.6<z<4.0$. They would set an incredible benchmark to compare with the low redshift data, and test exotic early Dark Energy models.

\begin{table}[htbp]
\centering
\begin{tabular}{cccc|ccc|c}
\hline\hline
Sample &$z$ & $\bar{n}$ & $V$ &$\sigma_{D_A}/D_A$ &  $\sigma_{H}/H$ & $\sigma_{D_V}/D_V$ &  $\sigma_{f\sigma_8}/f\sigma_8$ [\%] \\
 & & $[\rm10^{-4} h^3/Mpc^3]$ & $[\rm Gpc^3/h^3]$ & [\%]  & [\%] & [\%]   &  $k_{\rm max}=0.1 \rm[h/Mpc]$ \\
\hline
ELGs  & $1.6-2.0$ &$1.8$ &$15.56$& $0.81$ & $1.43$ & $0.56$ & $1.86$ \\
      & $2.0-2.4$ &$1.8$ &$16.20$& $0.74$ & $1.30$ & $0.51$ & $2.05$ \\
LBGs  & $2.4-2.8$ &$1.1$ &$16.27$& $0.96$ & $1.59$ & $0.64$ & $2.68$ \\
      & $2.8-3.2$ &$1.1$ &$16.00$& $0.94$ & $1.54$ & $0.63$ & $2.94$ \\
      & $3.2-3.6$ &$1.1$ &$15.54$& $0.93$ & $1.52$ & $0.62$ & $3.23$ \\
      & $3.6-4.0$ &$1.1$ &$14.99$& $0.94$ & $1.52$ & $0.62$ & $3.59$ \\
\hline\hline
\end{tabular}
\caption{Forecast constraints on BAO distance precision and growth of structure precision by MSE.}
\label{tab:baorsdresult}
\end{table}

For the RSD forecasts, we follow the Fisher matrix calculation in \cite{White2009}, and conservatively use modes within the scale range of $k < 0.1 \rm [h/Mpc]$. The RSD parameter can be constrained to $2.1$ per cent by MSE ELGs in two redshift bins. At higher redshifts, MSE LBGs can measure RSD precision at $3.6$ per cent level. These measurements rely on smaller scale observations than the BAO constraints and so the low density of LBGs in particular limits the precision achievable. However, the measurements will still test gravitational growth over a range of redshifts not previously probed in this way. Note that at $z=4$, for standard $\Lambda$CDM models, $f\sim0.99$, and we see that the RSD constraint is only weakly dependent on the gradient of the growth rate, and gives instead a strong measurement of $\sigma_8$. These measurements would therefore help to understand any remaining discrepancies between probes of structure growth, such as the current mismatch between weak lensing and CMB predictions.

\subsection{Discussion}
We have shown that an instrument such as MSE can answer two of the most important remaining questions within physics, namely determining the masses of neutrinos and providing insight into the physics of inflation. It will do this by targeting the high redshift Universe, measuring cosmological density fluctuations over an enormous volume - approximately 280\,Gpc$^3$. The large collecting area of MSE allows us to measure galaxy redshifts out to $z\sim4$ with exposure times that allow a large-area survey to be undertaken within a reasonable amount of time. The multiplexed spectroscopic capability matches that required to observe a population of galaxies with sufficient density to measure the large-scale overdensity modes required to understand Inflation. 

While we have shown that an exciting cosmological-focused survey is possible with the current baseline design, increasing the number of fibres would lead to improved accuracy in cosmological parameter measurements and would also provide margin for target selection. Increasing the number of fibres may only be possible with an increased FoV, but the increase of FoV itself is not important - it is the increase in the number of fibres provided that is. Moving MSE to the Southern hemisphere would allow the full LSST dataset to be used for targeting. However, we will have a sufficient photometric depth with CFIS + Union (+ Euclid) in the North to provide targets, so moving to the Southern hemisphere is not crucial for the science return. 

\begin{figure}
\begin{center}
\includegraphics[width=0.49\textwidth]{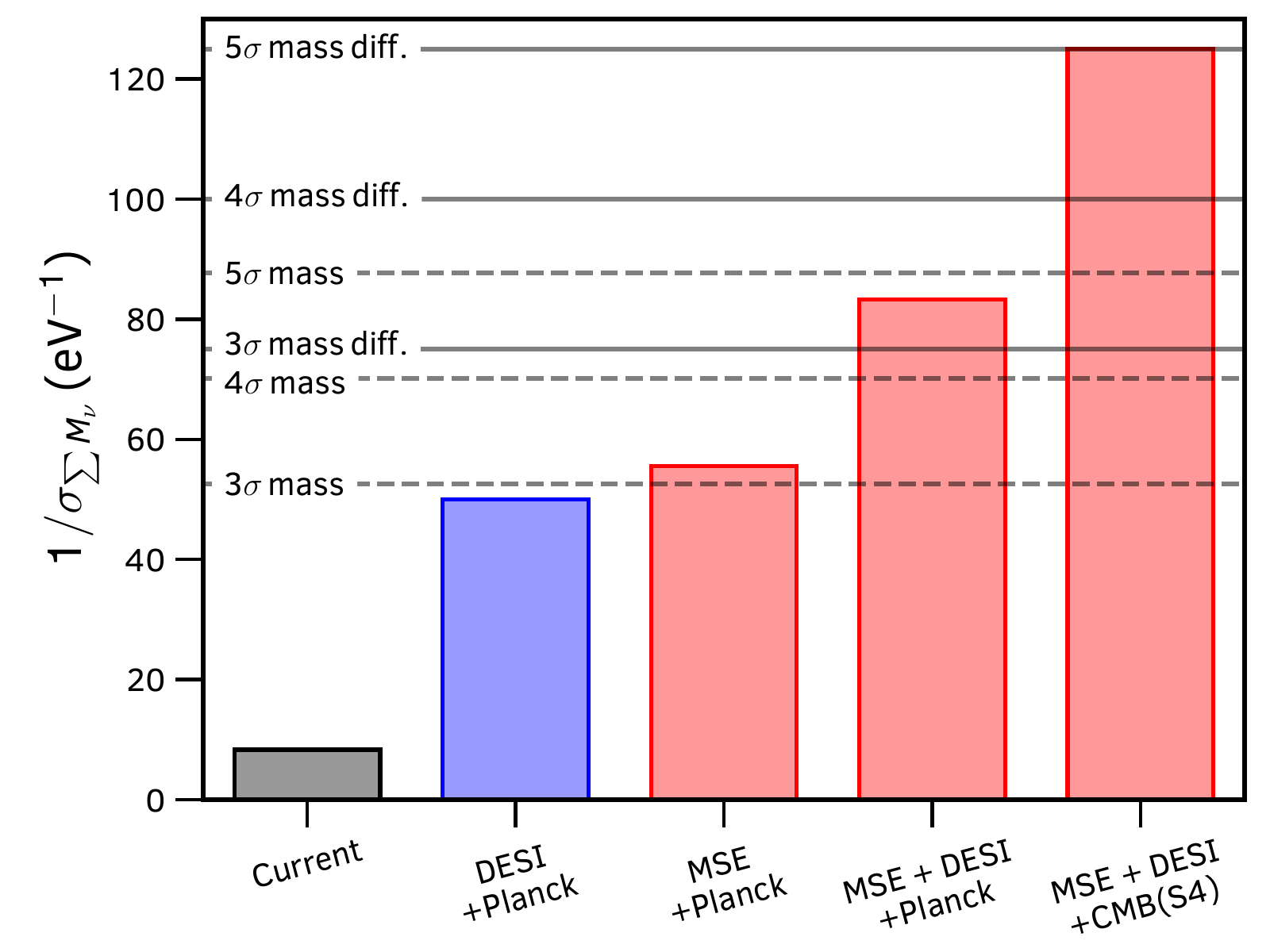}
\includegraphics[width=0.49\textwidth]{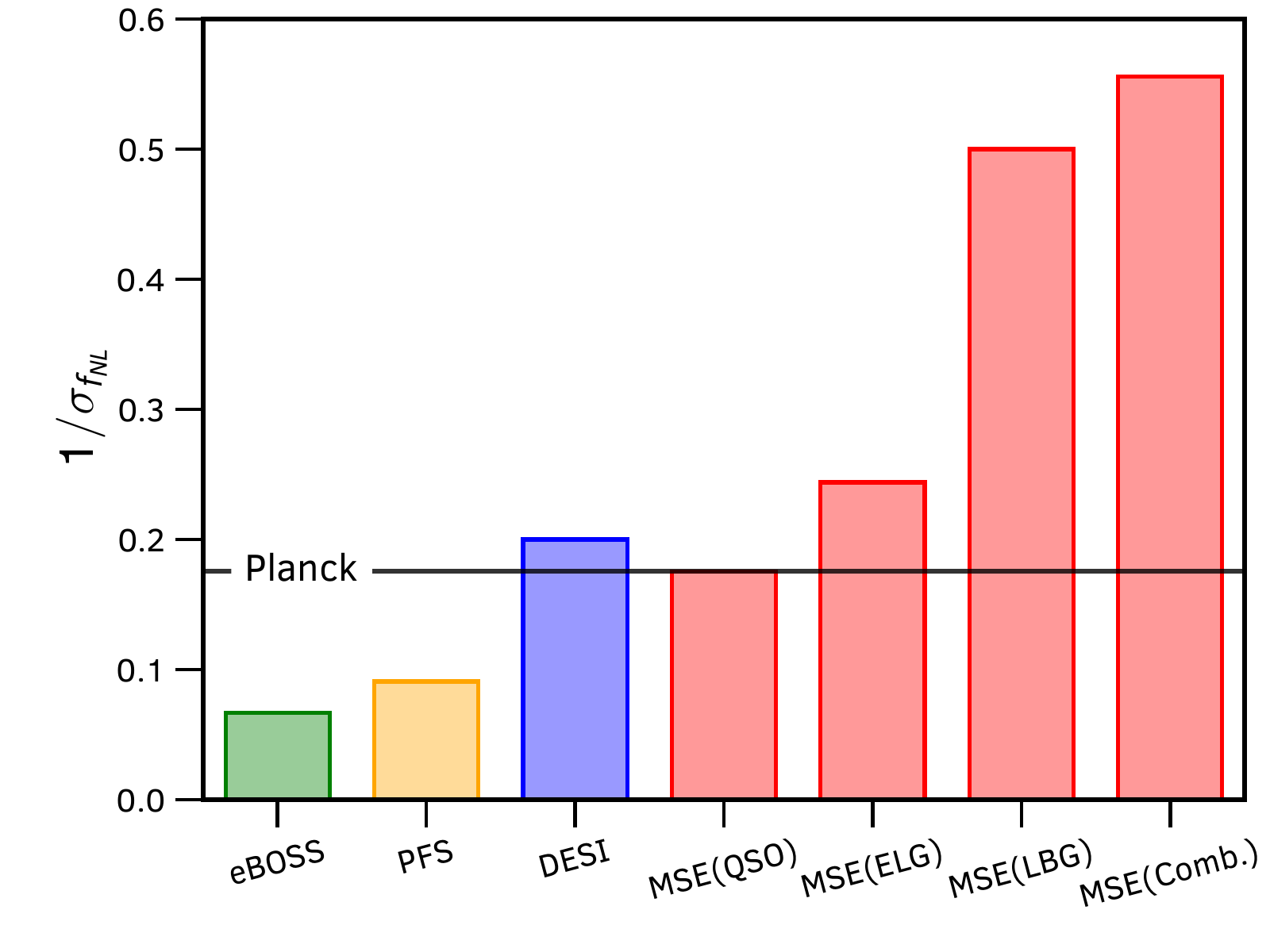}
\caption{A summary of the neutrino mass (\textit{left}) and primordial non-Gaussianity (\textit{right}) constraints achievable with the MSE compared to other surveys. Full details are given in the Sections~\ref{sec:neutrinos} and~\ref{sec:fnl}. The dashed and solid horizontal lines in the left panel show the requirements for $3$, $4$ and $5\sigma$ constraints on the sum of neutrino masses (either hierarchy) and on the mass difference between hierarchies respectively. The horizontal line on the right panel shows the current constraints on $f_{NL}$ from \textit{Planck} CMB data.}
\end{center}
  \label{fig:summary}
\end{figure}

The baseline survey clearly pushes beyond the capabilities of DESI and Euclid into a new regime for galaxy surveys. The focus of the predictions and design of the survey we have presented have been measuring neutrino mass and primordial non-Gaussianity, which we do with theoretically interesting precision. A summary of the improvement achievable by MSE over other projects is shown in Fig.~\ref{fig:summary}. We have also shown that the MSE High-z Cosmology Survey will provide unique BAO and RSD measurements over an untested redshift range, offering significant discovery space.

Our cosmological predictions focused on using galaxies as point-tracers of large-scale structure, as these measurements are the key driver of the survey design. The measurements made possible by observations of the ${\rm Ly}\alpha$ forest of absorption features in the spectra of high-z quasars, and their cross-correlation with the galaxy samples at the same redshifts are also very exciting, even though they only represent a small fraction of the targets. These observations represent excellent "value" in terms of the information content obtained about the large-scale structure for each observation.

Once obtained, a survey such as the MSE High-z Cosmology Survey will offer a goldmine of data to be mined for many different astrophysical applications, including providing insights into galaxy formation and evolution, cross-correlations science with the Cosmic Microwave Background, calibration of photometric redshift errors from other surveys, analyses of voids, clusters, filaments and other large-scale structures. The public release of the galaxy catalogues will provide the definitive high-redshift spectroscopic survey database, enabling serendipitous science similar to that enabled by the Sloan Digital Sky Survey (SDSS) public database, and will have broad impact in many areas, probably in entirely unanticipated ways, as has also been the experience with the SDSS.

\section{Other ideas for surveys with cosmological aims}

\subsection{A deep survey for LSST photometric redshift training}
All of the probes of dark energy to be employed by the Large Synoptic Survey Telescope (LSST) will rely on estimates of photometric redshifts for galaxies, either directly or indirectly. In simulations with perfect knowledge of galaxy spectral templates (or perfect spectroscopic training sets for machine-learning based algorithms) and expected LSST measurement errors, it is found that LSST is capable of delivering photometric redshifts with $2\%$ uncertainties; however, existing datasets of comparable depth yield uncertainties closer to $5\%$.  

Enabling optimal LSST photo-z's -- i.e., reducing uncertainties from 5\% to 2\% -- would require a better understanding of the relationship between galaxy colors and redshifts.  This requires spectroscopy of galaxies spanning the range of the properties of the photometric samples to be studied (e.g., objects with $i<25.3$ for the LSST weak lensing sample, going significantly fainter than the spectroscopic samples being obtained for Euclid training) with sufficient signal-to-noise ratio to achieve highly-secure redshift measurements for a large fraction of targets.  Such a sample would greatly improve LSST dark energy constraints, improving the figure of merit from baryon acoustic oscillations and weak lensing alone by almost $\sim 50\%$ (large-scale structure studies especially benefit due to the sharper maps provided by higher-fidelity photo-$z$'s). Cosmology with galaxy cluster counts and photometric supernovae (i.e., those with redshifts assigned based on their host photo-$z$ rather than spectroscopy) would likely benefit more.  

\cite{Newman2015} lays out a basic strategy for photometric redshift training spectroscopy for LSST. It is estimated that spectroscopy of 20-30,000 galaxies to the weak lensing magnitude limit of LSST ($i<25.3$) is necessary.  In order to limit the impact of sample/cosmic variance, it is optimal to distribute those objects over a number of widely-separated fields; 15 20-arcminute diameter fields (the minimum considered viable in \cite{Newman2015}, as smaller fields would not allow the characterization of galaxy clustering) would have similar variance to five widely-separated single MSE pointings, or one 10 square degree field.  Given that $\sim 3000$ objects can be targeted per MSE pointing, seven widely-separated pointings would be the minimum viable program; more pointings would enable better characterization of the impact of cosmic variance (estimates of standard deviations from seven samples are unstable) and the rejection of fields which are outliers at a particular redshift.  

This work is extremely synergistic with galaxy evolution observations of interest for MSE.  Many objects that would be targeted for galaxy evolution surveys would be relevant for photo-z training, so programs could be pursued simultaneously (purely targeting limited redshift ranges would be highly undesirable for this work, however).  Furthermore, photometric redshift training is fundamentally equivalent to the problem of determining the range of galaxy spectral energy distributions at a given luminosity and redshift; as a result, the proposed survey would provide strong constraints on models of galaxy evolution.  

Additionally, shrinking LSST photo-z's from $5\%$ per-object errors to $2\%$ will improve the sharpness of maps of the large-scale structure proportionally.  This will improve S/N in studies of relationships between galaxy properties and environment, measurements of galaxy clustering, and photometric identification of galaxy clusters and groups; LSST can help constrain the broader context of the galaxies whose spectra MSE would obtain.  Star-galaxy separation for studies of local dwarf galaxies and the Milky Way halo would also benefit from improved knowledge of galaxy SEDs provided by this sample.   For galaxy evolution studies, having more fields vs. fewer (at fixed area) improves measurements of galaxy clustering on the scales relevant for galaxy evolution studies, reduces sample/cosmic variance in count-based statistics (e.g., luminosity and mass functions), and enables variance to be well quantified. 

For photometric redshift training, in general we wish to obtain spectroscopic redshifts for as broad a range of galaxies within the magnitude range of LSST weak lensing and galaxy evolution samples. For weak lensing, this represents a limit of $i=25.3$; for galaxy evolution samples are currently less defined but going fainter than this spectroscopically for highly-complete samples is infeasible.  We then aim to obtain a signal-to-noise per angstrom equivalent to that obtained by the DEEP2 Galaxy Redshift Survey at $i=22.5$ (which yielded a $> 80\%$ redshift success rate at that magnitude) and comparable spectral resolution, but covering a broader wavelength range.  This leads to an estimated exposure time of 135 dark hours per pointing, for a total of 1000 (1500) hours for a 20,000 (30,000) object survey.  \footnote{Our estimate of 135 hours per pointing is based upon the scaling of photon statistics assuming the same throughput for MSE as for Keck/DEIMOS and that redshift success will be determined purely by SNR.  We have also estimated the time required to reach the same signal-to-noise ratio per angstrom as DEEP2 using the MSE exposure time calculator, and obtain results within a factor of two of this.  However, the ETC-based estimate is more uncertain, as the DEEP2 redshift success rate at $i=22.5$ can be measured directly from catalog-level DEEP2 data products, but signal-to-noise ratios for direct comparison to ETC results can only be determined from a new pixel-level analysis.  As a result, the ETC-based calculation relies on an uncertain (and spectrum-dependent) conversion from SNR at $r=23.5$ (as tabulated by DEEP2) to SNR at $i=22.5$, which is not an issue for the photon statistics conversion.}

If secure redshifts can be obtained for $> 99\%$ of the galaxies targeted for spectroscopy, this sample would provide a direct calibration of the redshift distributions of photo-$z$ samples accurate enough for LSST dark energy uncertainties to be dominated by random errors rather than photo-z systematics. However, given past redshift failure rates for deep samples of 20-40\%, it is quite likely that this high threshold will not be reached.  In that case, cross-correlations between LSST photo-z objects with galaxies with precision redshifts from wide-area spectroscopic surveys (such as the one proposed above) present the most promising route for this calibration, exploiting the fact that both sets of objects trace the same underlying large-scale-structure (Newman 2008). Even if $>99\%$ redshift success rates are not achieved, however, the proposed sample would reduce uncertainties on individual objects’ photo-z's from $\sim5\%$ to $\sim 2\%$ in those regions of parameter space with good redshift success, greatly enhancing the value of the LSST dataset for both galaxy evolution and cosmology studies.

We note that although we have focused on LSST here, it is worth noting that the proposed {\it WFIRST} satellite would have similar, overlapping but not identical spectroscopic training needs to enable cosmology with it.

\subsection{Pointed observations of galaxy clusters to z = 1}
Galaxy clusters may be detected using observational techniques sensitive to each main cluster component: weak-lensing is sensitive to the total cluster mass; the Sunyaev-Zeld'ovich (SZ) decrement and X-ray imaging are sensitive to the cluster ICM; galaxy overdensity studies are sensitive to the member galaxy population. MSE is poised to be an essential tool for cosmological studies based on galaxy clusters through high completeness targeted observations that will provide detailed information on their dynamical state for comparison to SZ, X-ray, and galaxy overdensity surveys. 

The landscape of galaxy cluster studies over the next two decades will be dominated by large samples of clusters identified using multiple, complementary techniques from observations executed over many thousands of square degrees. The aim of such studies is to obtain precise and accurate knowledge of our cosmological model, particularly the dark energy equation of state, in addition to compiling a detailed picture of how the mass and physical history of galaxy clusters in turn affects the evolutionary history of their member galaxies. There are two critical issues that affect the extent to which such galaxy cluster samples can be used to test cosmological models:
\begin{enumerate}
\item  The relationship between the observable used to identify each cluster and its true mass (in this case measured with respect to a common overdensity scale, e.g., M500,c).
\item  The relationship between a sample of clusters identified using a given observational method (e.g., SZ, X-ray, galaxy overdensity) and the true population of clusters existing in the universe.
\end{enumerate}

Cluster mass --- the key parameter whose evolution is predicted by theory and N-body simulations --- is a problematic parameter because it cannot be observed directly. Traditionally, cluster masses for large samples of clusters could only be inferred statistically, via various ''observable''-mass scaling relations. Popular observables include X-ray temperature/luminosity, SZ decrement, and cluster richness, ${\rm N_{gal}}$. However, there is currently no mass proxy that is simultaneously accurate (unbiased) and precise. Cluster cosmology benefits hugely from knowing the average cluster mass accurately (for the amplitude of the mass function) and the relative masses of clusters precisely (to get the shape of the mass function). For the absolute calibration, weak lensing currently achieves the highest accuracy. To get relative masses, X-ray gas mass and/or temperature are more useful because they have a smaller intrinsic scatter than WL.

So what can MSE bring to this field of study? Put simply, redshifts. Spectroscopic redshifts can be used to fix individual galaxies in space and thus determine cluster velocity distributions to a few percent, translating into a  $\sim 10\%$ accuracy in mass. This would provide a highly competitive mass calibration among the various cluster mass proxies. The statistical accuracy could be further improved by ''stacking'' clusters of similar values of observables (e.g., richness, X-ray temperature). MSE could also be employed to determine the recent star formation history of member galaxies viable the observation of narrow emission and absorption features generated within the photospheres of stars or in the interstellar gas.

\subsection{An IFU-based peculiar velocity survey}
In a peculiar velocity survey, one obtains an independent estimate of galaxy distance as well as the redshift. Although such surveys are necessarily limited to low redshift, this opens up tremendous scientific opportunities, competitive with high-redshift measurements aimed at probing the departures from General Relativity. Our proposed program will require an IFU survey of $\sim 300,000$ galaxies for which the distances will be independently estimated, using the Tully-Fisher and Fundamental Plane techniques, covering 20,000 square degrees, i.e. $\sfrac{3}{4}$ of the full sky apart from Milky Way. The survey will measure both early- and late-type galaxies up to redshift $z \sim 0.15$ using the Fundamental Plane and optical Tully-Fisher techniques. The scale is significantly greater than surveys that will be completed in 2025, such as the two Australian southern surveys: the optical multi-fiber TAIPAN ($\sim$ 50,000 galaxies) and the radio SKA Pathfinder WALLABY survey ($\sim$ 30,000 galaxies). This field of research is currently undergoing a revival, and this survey will achieve the same galaxy number density as in the leading contemporary and future surveys but over a much larger volume.

Two major science goals define this program: (i) linear-growth rate of cosmic structures at low redshift, obtained through velocity-density cross-correlations, which are free of cosmic variance. This allows one to discriminate between modified theories of gravity at $1\%$ level and yield a significant complement to the high-redshift constraints provided by the coeval spectroscopic surveys achieved e.g., by Subaru/PFS, Mayall/DESI, Euclid, and WFIRST;  (ii) dynamical tests probing the scale of homogeneity.

Having independent distance and velocity measurements means that we can measure peculiar velocities directly (as opposed to inferring them from the quadrupole of the power spectrum as in RSD). The comparison of the peculiar velocity field with the density field then eliminates cosmic variance. This comparison is at the heart of peculiar velocity studies \citep{Davis2011}. As shown by \cite{Carrick2015}, the measurement of the normalized linear growth-rate $f\sigma_8$, based on $\sim 5,000$ peculiar velocities with a typical depth of $z \sim0.02$ has a precision of $5\%$; with the proposed peculiar velocity from MSE, one might expect to reduce this uncertainty by a factor of $\sim 5$, yielding a precision of $\sim 1\%$. In addition to not being affected by cosmic variance, peculiar velocity surveys are also the ideal framework to investigate primordial non-Gaussianities and general-relativistic (gauge) effects specific of large scales \citep{Jeong2012,Villa2014}, and to test the scale-dependence of the growth rate of structure, a feature prevalent in modified theories of gravity and clustering dark-energy models \citep{Parfrey2011}, see also the review by \cite{Clifton2012}.

Of course, such a large IFU galaxy survey would also be extremely valuable for a range of galaxy science.

\bibliography{biblio}

\end{document}